\documentclass[showpacs,preprintnumbers,amsmath,amssymb,twocolumn,superscriptaddress,aps,pre]{revtex4}

\usepackage{ifpdf}
\ifpdf	
	\newcommand{\ignoreThis}[1]{}
	\usepackage{color}	
	\definecolor{Gray}{rgb}{0.6,0,0}
\else		
	\newcommand{\ignoreThis}[1]{#1}
	\usepackage[usenames,dvips]{color}
\fi 
\usepackage[normalem]{ulem}

\usepackage{epsfig,graphics,graphicx,amssymb}
\usepackage{multirow}
\usepackage{framed}
\usepackage{xr}
\usepackage{url}

\begin{document}

\title{Simplification and analysis of a model of social interaction in voting}
\author{Luis F. Lafuerza}
\affiliation{Theoretical Physics Division, School of Physics and Astronomy, The University of Manchester, Manchester M13 9PL, UK}
\author{Louise Dyson}
\affiliation{Theoretical Physics Division, School of Physics and Astronomy, The University of Manchester, Manchester M13 9PL, UK}
\affiliation{Current address: Mathematics Institute, University of Warwick, Coventry CV4 7AL, UK}
\author{Bruce Edmonds}
\affiliation{Centre for Policy Modelling, Manchester Metropolitan University, Manchester, M15 6BH, UK}
\author{Alan J. McKane}
\affiliation{Theoretical Physics Division, School of Physics and Astronomy, The University of Manchester, Manchester M13 9PL, UK}



\begin{abstract}
A recently proposed model of social interaction in voting is investigated by simplifying it down into a version that is more analytically tractable and which allows a mathematical analysis to be performed. This analysis clarifies the interplay of the different elements present in the system --- social influence, heterogeneity and noise --- and leads to a better understanding of its properties. The origin of a regime of bistability is identified. The insight gained in this way gives further intuition into the behaviour of the original model. 
\end{abstract}

\maketitle

\medskip

{\large\textbf{1\ Introduction}}

\medskip

There is a growing appreciation of the importance of social influence in voting \cite{Zuckerman05,Rolfe12,Sinclair12}, and convincing experimental evidence of the phenomenon has recently been produced \cite{Nickerson08,FBvoting12}. However, a detailed understanding of the process and its implications is still lacking. Systems of interacting elements can display complex, sometimes counter-intuitive, behaviour \cite{complexity}, making mathematical analysis highly useful to understand the properties of such systems.

There are already a number of studies modelling voting as a social influence process, for example \cite{Sznajd1,Sznajd2,Borghesi1,Borghesi2,Galam91,Fowler,Anxovoting11,Voterdata}. These tend to consider rather simple models that intend to capture, in a stylised manner, some aspects of the voting process or to reproduce some observed regularity. A different approach was taken in Refs.~\cite{modelref,modelref2}, where a collaboration between social scientists and computational modellers led to the creation of a complex computational model of voter turnout. There is a tendency for the former approach to be taken by physicists, who have a tradition of gaining intuition through the use of simple models in their own subject. On the other hand, the latter approach is frequently favoured by social scientists, who wish to include all aspects which they feel may have an influence on the system. This difference in approach has the unfortunate consequence of leading to the formation of two groups of modellers whose models have little in common, and who have little incentive to communicate.

In a previous paper~\cite{SCIDVoterUS}, we started to address this issue by forming a bridge between the two methodologies. This consisted of constructing an intermediate model which was between the two types described above. The philosophy behind this is discussed in some detail in Ref.~\cite{SCIDVoterUS}, but we had several goals in mind. One was simply to attempt to bring together the two communities described above, by formulating a model which had features of both perspectives. Another was to develop the methodology of forming such intermediate models. The actual procedure we adopted in constructing the new model consisted, very broadly, of beginning from the complex model of Refs.~\cite{modelref,modelref2}, and systematically eliminating certain features which did not have a marked effect on the outcomes of simulations. 

We would not necessarily expect that a single intermediate model would be able to bridge the large gap between complex models and the models favoured by physicists, and therefore we instead envisage there being a sequence of intermediate models, each less complicated than the previous one, while retaining sufficient features in common with its `neighbours' in the model sequence, that any similarities and differences between them can be systematically studied and the reasons behind these understood. In the context of models of voter turnout which interest us here, we will denote the most complex model of Refs.~\cite{modelref,modelref2} as Model 1 and the simplified version of this model discussed in Ref.~\cite{SCIDVoterUS} as Model 2. The purpose of this paper is to create a further model in the sequence, denoted as Model 3, which comes near to being a model of the type preferred by physicists, in that it is sufficiently simple to allow some mathematical analysis.

The method used to reduce the model complexity also uncovered a number of phenomena which were difficult to detect in Model 1 (simply because of its sheer complexity), and revealed several mechanisms required for these phenomena to be observed. One such phenomenon was the existence of a single control parameter (the `influence rate') that largely controlled the levels of voter turnout in the model. When the influence rate was low, simulations of Model 2 always resulted in low voter turnout. Conversely for large influence rates, voter turnout was always high. For intermediate values of the influence rate, the reduced model displayed bistability, so that different runs of the same model with the same parameters and initial conditions could, by chance, give either a high or a low turnout. The construction of the further simplified Model 3 should allow us to gain a deeper understanding of this phenomenon.

The outline of the paper is as follows. In Section 2, we first give an overview of the differences between the previous two models (more detail is given in the Appendices), describe the new model (Model 3) and then give the results of simulation which show that the predictions of Model 2 and Model 3 are qualitatively similar. Having established this essential requirement of Model 3, we go on to mathematically analyse it in Section 3. We conclude in Section 4 with a summary and a look to the future. 

\medskip

{\large\textbf{2\ Formulation}}

\medskip

\begin{figure}[ht]
 \centering
 \includegraphics[width=0.45\textwidth]{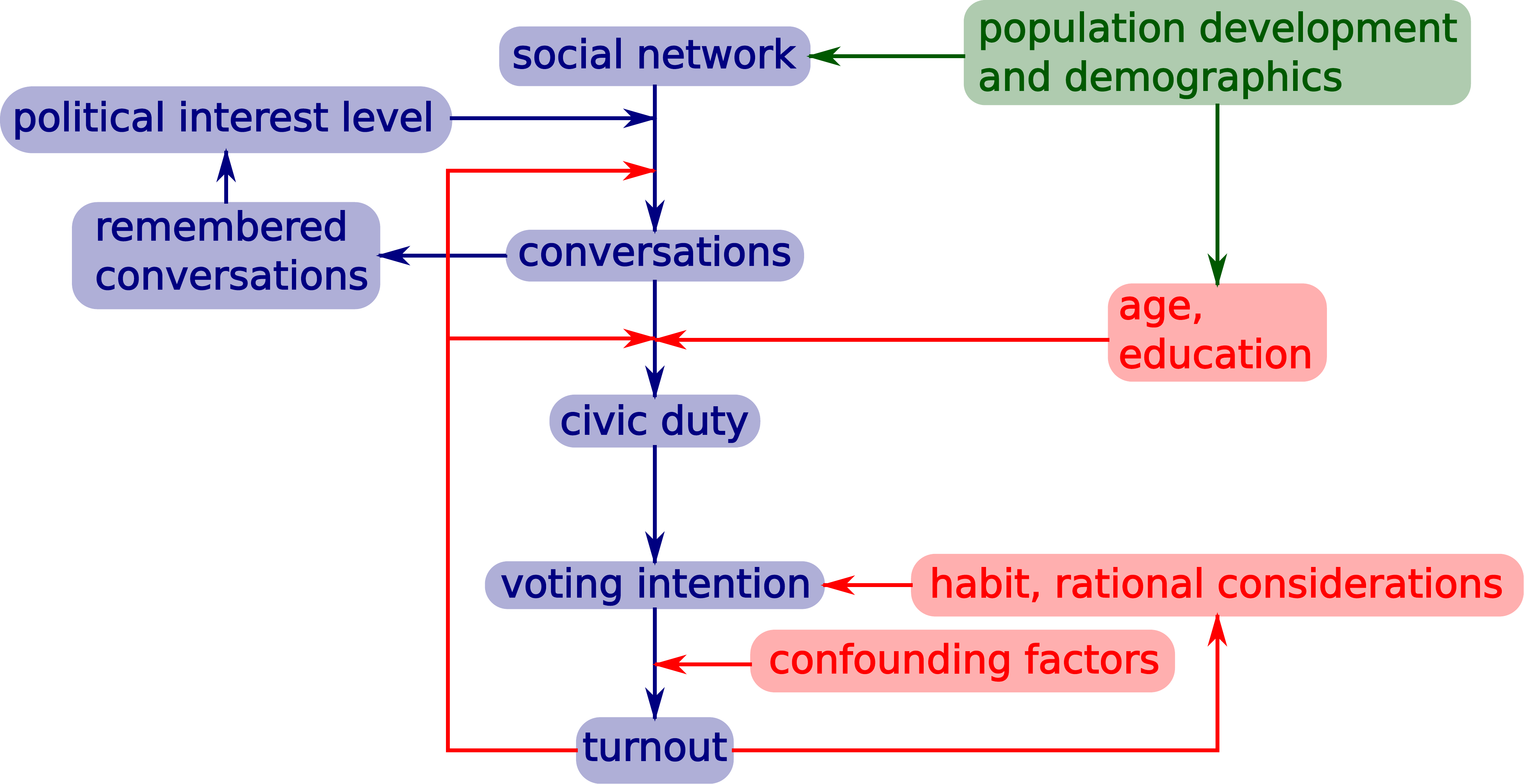}
 \caption{\textbf{Diagrammatic representation of the complex model.} The main pathway is shown in blue, with additional factors in red, and general development of the agent population in green.}\label{F:OM}
 \end{figure}

Model 1 is a complex agent-based computational model of voting, developed to incorporate the evidence suggested in the social science literature. This model describes the inhabitants of a neighbourhood or small city, and consists of a population of agents that occupy the sites of a square lattice, with sites corresponding to houses, workplaces, schools and other places of activity. The agents have a large number of characteristics (including age, ethnicity, interest in politics and `civic duty') and are subjected to many processes (including ageing, moving house, finding jobs and having children). These processes modify some of the agents' characteristics and allow them to make links to other agents, creating a social network. Agents also initiate (political) conversations over this social network, with a probability that depends on their political interest. In turn, the conversations they receive affect their political interest and allow civic duty to spread between agents. Agents' civic duty (together with other elements) determines their propensity to vote in a series of periodic elections. Agents can leave the simulation by dying or emigrating and new agents are created via births and immigration. The details of the model can be found in \cite{modelref,modelref2}. A schematic representation of the model is given in Fig.~\ref{F:OM}.


In a recent work, Model 1 was distilled into the more compact Model 2. By neglecting some of the more complicated components of Model 1, Model 2 allowed us to investigate the effects of different mechanisms, and to explore the parameter space in a more efficient way. A major simplification of Model 2 was ignoring  all processes that form links between agents, setting instead a social network with appropriate characteristics. Other elements, such as those regarding party preference and development of children, were also neglected. Model 2 achieved a large simplification (of the order of a factor $1000$ in computational efficiency \cite{SCIDVoterUS}) whilst maintaining good agreement with Model 1. A detailed description of the relatively complex Model 2 is given in Appendix~\ref{appendixmodel2}.

In this section we will push the model-analysis process forward. We will start by formulating Model 3, a simpler version of Model 2 that is more suitable for mathematical analysis. Some of the simplifications we will make will involve formulating processes in a more standard form so that checking the output of these against those from Model 2 enables us to draw conclusions that are robust with respect to the implementation details.  

In Model 2, social influence (the key mechanism in the model) is implemented using two main variables: interest level and number of conversations remembered. The interest level determines the propensity of an agent to initiate conversations. In turn, the interest level of an agent is determined by the number of their received conversations (together with their minimum interest level). Conversations are forgotten with some probability every year. A further complication is that agents with zero interest level have different dynamics, until their number of `background' conversations remembered exceeds a given threshold (see Appendix~\ref{appendixmodel2} for details). 

We will simplify Model 2 in three main ways to create Model 3. Firstly we will use the same dynamics for all interest levels (thus ignoring the difference in Model 2 for agents with interest level zero). In addition, we will use a single variable for the influence process, which we call the `interest state', that increases with each received conversation and decreases appropriately, rather than one variable for interest level and a different one for conversations remembered. Moreover, we will uncouple the dynamics of the interest state from that of voting. While in Model 2 agent's voting behaviour affects their probability to initiate conversations (turnout-conversations feedback in Fig. \ref{F:OM}), this feedback will be neglected in Model 3.
 
This formulation is more suitable for mathematical analysis and will allow us to see the interplay of the different elements of the model in a more transparent way. We define Model 3 below.

\medskip

{\large\textbf{2.1 \ Model 3 definition}}

\medskip

The model consists of a population of $N$ agents. Agents enter and leave the model via immigration-emigration and birth-death. For simplicity we keep the population size constant, by matching each death event by a birth and every emigration by an immigration. Agents age throughout the simulation and their age determines their death probability. The $i$th agent is characterized by three dynamic variables: (political) interest state, $s(i)$; civic duty, $d(i)$; and voting habit, $h(i)$. In addition, agents have two fixed characteristics: their `intrinsic interest' state, $m(i)$; and their level of education, $ed(i)$. The interest state, $s(i)$, is the primary dynamic variable in the model, controlling how often the agent initiates conversations, and depending on the number of received conversations. Civic duty, $d(i)$, (spread through conversations) and voting habit, $h(i)$, are binary variables that together determine the probability of an agent voting. Finally, the intrinsic interest state determines the minimum value that the interest state variable may take for that agent, and the education modulates the probabilities of acquiring or losing civic duty. 

We now describe the model dynamics. 
\begin{itemize}
 \item[(i)] At each time step each agent initiates $Bin(K,f(s(i),m(i)))$ conversations~\footnote{If $f(s(i),m(i))>1$ then $K + Bin(K,f(s(i),m(i))-\lfloor f(s(i),m(i))\rfloor)$ conversations are realised, where $ \lfloor x\rfloor$ denotes the integer part of $x$. If $K$ is not a natural number, then $Bin(\lfloor K\rfloor,f(s(i),m(i))) + Bin(1,(K-\lfloor K\rfloor) f(s(i),m(i)))$.}, where $Bin(N,p)$ is a Binomial random variable for the number of successes from $N$ trials, each with probability $p$ of success. 
\item[(ii)] For each successful conversation another agent, $j$, is picked at random (corresponding to a fully connected underlying network) and the receiving agent increases their interest state [$s(j) \rightarrow s(j)+1$]. Conversations can lead to the spread of civic duty in the following way. If a conversation takes place from an agent, $i$, with civic duty to an agent, $j$, without it, and agent $j$'s interest state is larger than some threshold, $s(j)\geq T_d$, then they gain civic duty with probability, $a_d$, dependent on whether they voted in the last election.
\item[(iii)] Each time step, after all conversations are completed, every agent has a probability $[s(i)-m(i)]\gamma$ to decrease their interest state by one. Thus in the absence of received conversations an agent's interest state will decay to their intrinsic interest state over a time-scale of order $1/\gamma$. In addition, agents lose civic duty with a probability $l_d$ so that civic duty decays over a time-scale of order $1/l_d$ (typically $l_d \ll \gamma$).
\item[(iv)] Elections take place with periodicity $\tau_e$. An agent will vote with probability (1-$p_c$) if they have civic duty or voting habit, and will otherwise not vote. Here $p_c$ is the probability of not voting, despite having the intention to vote, due to ``confounding factors'', for instance illness or having recently lost employment. An agent voting in three consecutive elections acquires voting habit, and loses this habit if they fail to vote in two consecutive elections.
\end{itemize}
We focus mainly on a fully connected underlying network to ease the analysis. By initially ignoring network effects we can better understand the main properties of the model. Network effects are considered at the end of Section 3.
\medskip

{\large\textbf{2.2\ Comparing Model 2 and Model 3}}

\medskip


\begin{figure}[h]
  \centering
\includegraphics[angle=0,scale=0.3]{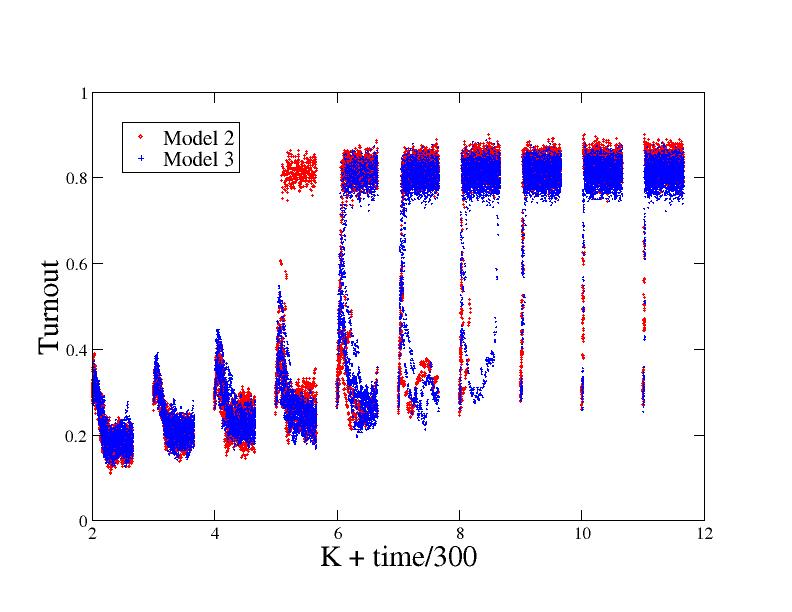}
 \includegraphics[angle=0,scale=0.3]{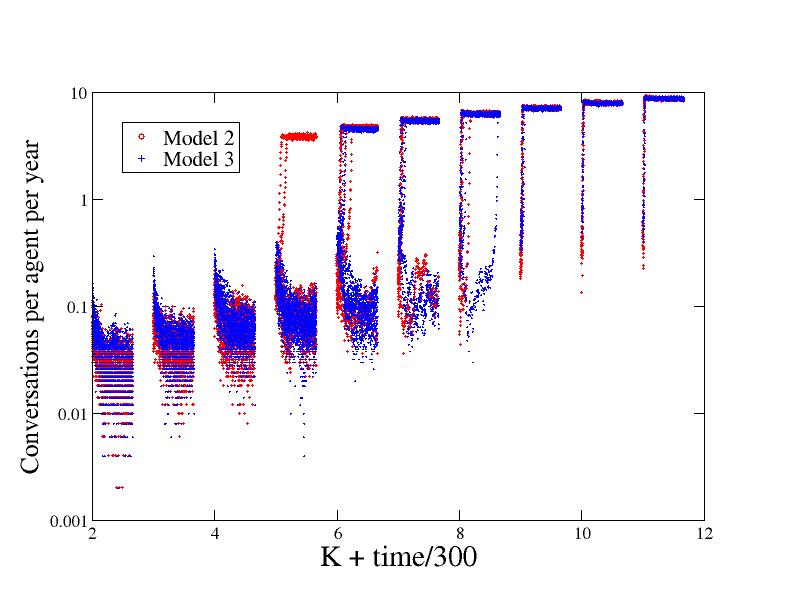}
   \caption{Comparing the results for Model 2 (red circles) and Model 3 (blue pluses), in turnout (top figure) and number of conversations per year per agent (bottom, in log-scale to appreciate the low-conversations regime better). For each value of $K$ (the influence rate), we plot $25$ time series from $0$ to $200$ years. Here $c=3$ and other parameter values are given in Appendix~\ref{appendixparam}.} \label{fig:comparing}
\end{figure}


Model 2 (with an underlying fully connected network) and Model 3 outputs are compared in Fig.~\ref{fig:comparing}. For each value of the influence rate, $K$, we show $25$ different time series from $0$ to $200$ years, corresponding to different realisations of the process. Due to the non-standard nature of this figure, we stress again that each of the ten collection of points that is seen in the figures represent time series running from $0$ to $200$ years. We can see that there are essentially two regimes, one with a small number of conversations and low turnout, obtained for low values of influence rate, and another with a large number of conversations and high turnout, for higher values of the influence rate. These two regimes are connected by a region of bistability, in which, for the same parameter values and initial conditions, some realisations converge to the `high-communication' regime and some to the `low-communication' one (which regime is achieved is a random outcome arising from the stochasticity of the process). We note that the results are similar when plotting the variables civic duty or voting habit.

We can also see from Fig.~\ref{fig:comparing} that results of Model 3 and Model 2 show good agreement. The number of conversations are very close for the two models. The main difference between the two models is that Model 3 gives a slightly smaller voter turnout. In the low-communication regime, this is primarily due to the simplifying assumption in Model 3 that individuals with zero interest state follow the same dynamics as those with higher interest states (which is not the case in Model 2). This leads to a smaller number of agents being susceptible to acquiring civic duty. In the high-communication regime, the difference in turnout is mainly due to the assumption in Model 3 that the probability, $p_c$, of an agent not voting in spite of having civic duty or voting habit, does not depend on age. In contrast in Model 2, the probability of an agent with civic duty or voting habit not voting is smaller for younger agents, so that agents in Model 2 are slightly more likely to build habit, leading to higher levels of voting (the aforementioned zero-interest effect is less important in this regime because almost no agent has interest state equal to zero).  This explanation is confirmed by simulations which include these elements. Despite these small quantitative differences, Model 2 and Model 3 are qualitatively very similar.

\medskip

{\large\textbf{3\ Analysis}}

\medskip

We would like to understand the origin of the low- and high-communication regimes and the bistability region, as well as the mechanisms required to observe these features. In order to do so, we will perform a general mathematical analysis of Model 3. Because the interest state dynamics is unaffected by the voting dynamics and yet interest and voting are closely correlated, we conclude that the interest state dynamics is the main driver of the system, and will focus our analysis on the interest state dynamics. 

The process is a discrete-time analogue of the following system of continuous-time birth and death stochastic processes:
\begin{equation}
 s(i) \overset{\beta}{\longrightarrow} s(i)+1, \hspace{1cm}s(i) \overset{\delta_i}{\longrightarrow} s(i)-1,\label{schema}
\end{equation}
with $\beta\equiv K\sum_{j}f(s(j),m(j))/N$ and $\delta_i\equiv [s(i)-m(i)]\gamma$ (note that if $s(i)\geq m(i)$ initially then $\delta_i$ is always positive). We will analyse this continuous-time version for mathematical convenience.

In order to make analytical progress, we will assume that $\beta$ is time-independent. This is suggested by applying the central limit theorem, since $\beta$ is the sum of $N$ independent random variables divided by $N$ (the central limit theorem does not strictly apply here, since the $s(j)$ variables are not independent, so the time-independence of $\beta$ is an assumption), and it is supported by numerical simulations with large $N$. With the constant $\beta$ assumption, (\ref{schema}) implies that the variables $s(i)-m(i)$ follow independent linear birth and death processes. At steady state, we have \cite{VanKampenBook}:
\begin{equation}
 s(i)=m(i)+\text{Poisson}_i(\lambda),\label{Poissons}
\end{equation}
where $\lambda\equiv\beta/\gamma$, and $\text{Poisson}_j(\lambda)$ are Poisson random variables with mean $\lambda$ (independent for different values of $j$). 

Under the constant $\lambda$ assumption, inserting Eq. (\ref{Poissons}) in the definition of $\lambda$, we obtain the following self-consistent equation:
\begin{equation}
 \lambda=\frac{K}{\gamma}\sum_{j}\frac{f(m(j)+\text{Poisson}_j(\lambda),m(j))}{N}\label{self-consgen}.
\end{equation}
We can re-write the sum in (\ref{self-consgen}) grouping agents with the same value of the minimum interest state, $m$:
\begin{eqnarray}
 &&\sum_{j=1}^Nf(m(j)+\text{Poisson}_j(\lambda),m(j))=\nonumber\\
&&\sum_m\sum_{\alpha\in A_m}f(m+\text{Poisson}_{\alpha}(\lambda),m)\label{poppoiss},
\end{eqnarray}
with $A_m$ equal to the set of indices of agents that have intrinsic state equal to $m$, $A_m=\{j|m(j)=m\}$. Assuming that $A_m$ is large for every $m$ (\emph{i.e.} there are many agents of each type) or, equivalently, disregarding fluctuations, $\sum_{\alpha\in A_m}f(m+\text{Poisson}_{\alpha}(\lambda))$ is just an average over a Poisson distribution with mean $\lambda$, 
\begin{equation}
 \sum_{\alpha\in A_m}f(m+\text{Poisson}_{\alpha}(\lambda),m)\approx|A_m|\langle f(m+n,m);\lambda\rangle,
\end{equation}
where $\langle;\lambda\rangle$ indicates an average over a Poisson distribution with mean $\lambda$, and $|A_m|$ denotes the number of elements in the set $A_m$, $|A_m|=NP(m)$, with $P(m)$ the fraction of agents with intrinsic interest state equal to $m$.
Making this approximation, Eq.~(\ref{self-consgen}) leads to:
\begin{equation}
 \frac{\gamma}{K}\lambda=\sum_mP(m)\langle f(m+n,m);\lambda\rangle\equiv g(\lambda).\label{self-consnice}
\end{equation}
Equation (\ref{self-consnice}) displays the key elements in the system and it is the main result of this section. The social interaction appears through the function $f$, the heterogeneity in the population via the average over the distribution of $m$, $P(m)$, and the intrinsic stochasticity through the average over the Poisson distribution. The form of the equation depends on the mean-field-type of interaction (equivalently, fully connected social network) assumed. The equation shows how the interaction function, $f$, smoothed out by the heterogeneity and stochasticity, determines the number and type of solutions as $K/\gamma$ is varied.

\begin{figure}[h]
  \centering
\includegraphics[angle=0,scale=0.5]{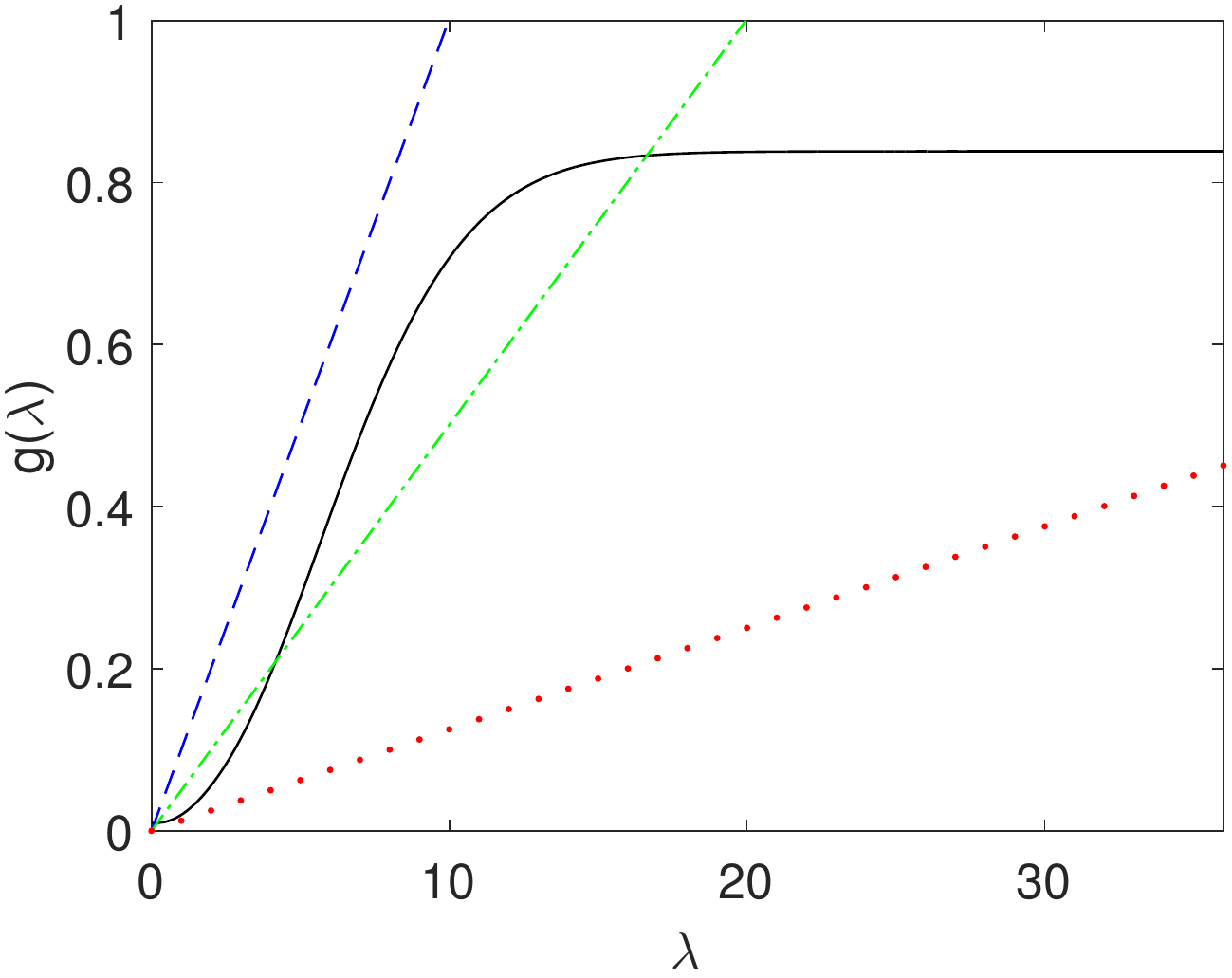}
 \caption{Right-hand side of Eq.~(\ref{self-consnice}) (solid line) and left-hand side for $K=2$ (dashed), $K=4$ (dot-dashed) and $K=16$ (dotted), for $c=3$ and other parameter values as in Appendix~\ref{appendixparam}. Between $K=2.8$ and $K=10.4$ the self-consistent equation (\ref{self-consnice}) has three solutions.} \label{self-consfig}
\end{figure}

In order to gain some intuition into the properties of Eq.~(\ref{self-consnice}), we will analyse it in the particular case in which $f(s,m)$ has the form used in Model 2 (see Appendix~\ref{appendixmodel2}). The formula for $g(\lambda$) corresponding to this case is given in Appendix~\ref{appendixparam}, and it is illustrated in Fig.~\ref{self-consfig}.


We see that $g(\lambda)$ displays an S-shape, starting at a small value for $\lambda=0$, increasing for intermediate values of $\lambda$ and saturating for larger values of $\lambda$. The value of $g(\lambda)$ at $\lambda=0$ is related to the amount of conversations that take place even when social influence is absent (due to agents with large intrinsic interest), while the value of $g(\lambda)$ for large $\lambda$ corresponds to the amount of conversations when all the agents have their maximum possible interest (due to large social influence). The position and sharpness of the increase depends on the form of $f$ as well as on the distribution of $m$. The sharper $f$ is, and the more homogeneous the population, the sharper the increase in $g(\lambda)$. We see that bistability is possible if $g(\lambda)$ increases relatively sharply, which corresponds to a sharply increasing $f$ and a homogeneous population, and if $\lim_{\lambda\rightarrow\infty}g(\lambda)-g(0)$ is large (compared with $g(0)$), which corresponds to the case in which social influence has a strong impact on the overall conversation levels. This prediction is confirmed by numerical simulations of Model 3, as well as of Model 2 and Model 1, illustrating how the analysis of the simpler Model 3 can generate important insights into the behaviour of the more complex Model 1.

\begin{figure}[h]
  \centering
 \includegraphics[angle=0,scale=0.3]{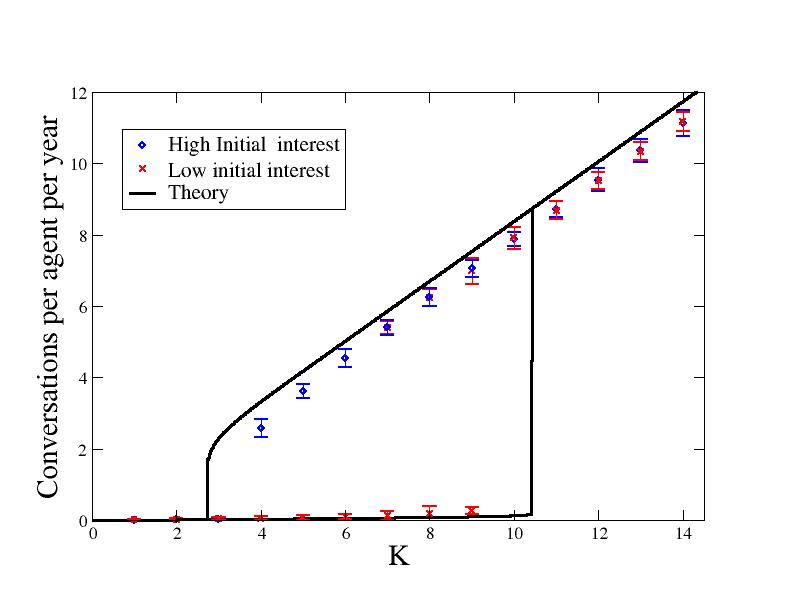}
 \caption{Number of conversations per agent per year as a function of parameter $K$, for $c=3$ and year $t=200$. Error bars correspond to two standard deviations. The red crosses correspond to runs in which the initial population had a low interest state while the blue diamonds correspond to runs with high initial interest. The simulations show a region of bistability between $K=4$ and $K=9$.} \label{convsfig}
\end{figure}

In Fig.~\ref{convsfig} the solutions of Eq. (\ref{self-consnice}) are compared with numerical simulations, showing good agreement.


\begin{figure}[h]
  \centering
  \includegraphics[angle=0,scale=0.3]{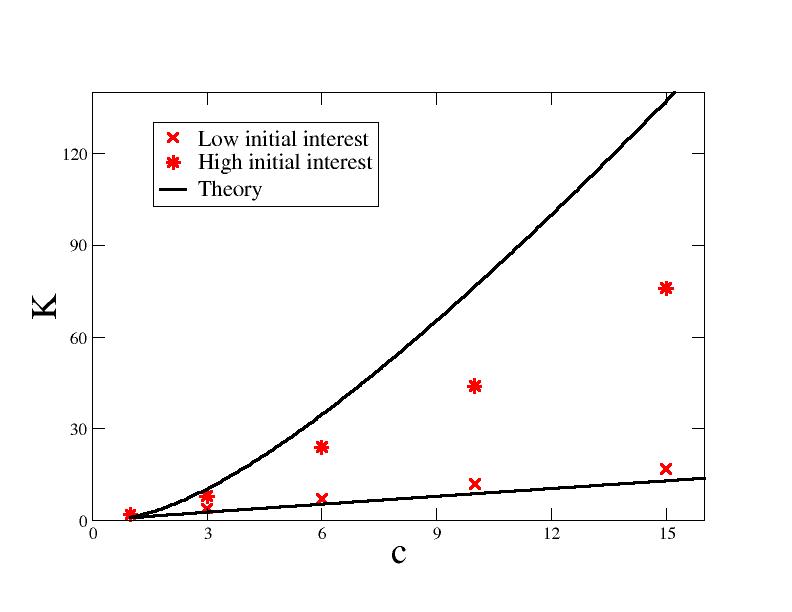}
 \caption{Bistability region in the $K-c$ plane. Bistability is observed for values of $K$ between the symbols. Theoretical predictions are displayed as solid lines.} \label{bifurcation-fig}
\end{figure}

Fig.~\ref{self-consfig} illustrates how for large values of $\gamma/K$ there is only one solution of Eq.~(\ref{self-consnice}). As $\gamma/K$ decreases, the line $(\lambda/K)\gamma$ becomes tangent to the right-hand side of (\ref{self-consnice}), giving rise to two new solutions through a saddle-node bifurcation. As $\gamma/K$ decreases further, a new tangent condition is obtained, leading to the disappearance of two of the solutions through another saddle-node bifurcation. Imposing Eq.~(\ref{self-consnice}) together with the equality of the derivatives leads to:
\begin{eqnarray}
 \lambda&=&\frac{g(\lambda)}{g'(\lambda)},\label{lambdacrit}\\
K&=&\frac{\gamma}{g'(\lambda)}.\label{Kcrit}
\end{eqnarray}
Solving Eq.~(\ref{lambdacrit}) and using Eq.~(\ref{Kcrit}) we can derive the values of $K$ and $\lambda$ for which the bifurcations take place, delimiting the region of parameter space where bistability is possible. The results of this exercise are presented in Fig.~\ref{bifurcation-fig}, together with results coming from numerical simulations of the model. The $x$-axis of this figure ($c$) is a parameter that determines the shape of $g(\lambda)$ (see Appendix B for its definition). The simulation-based results corresponding to the lower transition were determined as the minimum value of $K$ for which some realisations reach the high-communication regime, taking as initial condition a population with the maximum interest state; the results corresponding to the upper transition were determined as the maximum value of $K$ for which some realisations stay in the low-communication regime, taking as initial condition a population with the minimum interest state. We see that, for reasonably large values of $c$, the bistability disappears much sooner in the simulations than in the theoretical predictions. This might be due to the fact that, while, in the deterministic limit, the solution corresponding to the low-communication regime still exists, its relative stability  (and its basin of attraction) is low and the noise pushes the system to the `high-communication' solution. This particularly simple bifurcation diagram is obtained when the function $f$ used is taken to approximate Model 2 (see Appendix B). For more general forms of the function $f$ a more complex bifurcation diagram is possible.


It is interesting to compare Model 3 with other models of social influence. Threshold models of collective behaviour \cite{Granovetter78,Siegel09} form probably the simplest class of models showing similar phenomenology. In these models, an agent participates in an activity if some proportion of the population (the agent's threshold) also does so. Bistability can appear when the population is homogeneous (the threshold distribution is sharply peaked around a given value). Our model is similar to this class in that different agents require different amounts of social input in order to show the same level of activity (due to different intrinsic interests, $m$). The random-field Ising model \cite{randomIsing}, which can also be used to model social phenomena \cite{Galam91,Bouchaud05, BouchaudReview}, shows similar phenomenology, with multi-stability arising for small disorder and low noise. In this case, the local fields of the random-field Ising model would correspond to intrinsic interest of our model. The key ingredients of all these models are a heterogeneous population and a nonlinear influence function.

So far we have assumed a fully connected interaction network. The results can, however, change qualitatively when a different type of interaction network is used. The effects of the network in Model 3 are the same as those obtained for Model 2 \cite{SCIDVoterUS}. The main result is that the bistability region can disappear, with the relevant variables increasing in a smoother way, when the interaction network becomes sparse or strongly clustered. This effect is reminiscent of that found in random field Ising models in which sparser networks tend to require higher interaction strength for bistability to occur \cite{BouchaudReview}. From the perspective of our analysis, adding a network leads to conversation inputs that fluctuate more between agents. This results in an extra source of heterogeneity in the equation for $\lambda$ (that in this case should be replaced by $\overline{\lambda}\equiv\sum\lambda_i/N$) that further smooths the interaction function $f$, which results in a smaller region of parameter space in which bistability occurs. We note that since we expect real systems to be heterogeneous this result suggest that bistability may not be widespread in real-world systems.

\medskip

{\large\textbf{4\ Conclusions}}

\medskip

In this paper we have demonstrated the power of a method of model reduction involving generating a chain of increasingly simpler models. Beginning from a complex model (Model 1), and a previously published reduced model (Model 2), we have created a further reduced model (Model 3) and shown that it agrees quantitatively and qualitatively with Model 2. Since the relationship of Model 2 to Model 1 has been previously investigated, we thus have a clear link between results derived and understood in Model 3 to equivalent effects found in the original complex model, which in turn explicitly models mechanisms believed to be important in real world voting processes. 

We have described some advantages of this approach in the Introduction, but one of the most compelling is that it combines the best of two worlds: the simplicity appreciated by those trained in the physical sciences, but having an input from the many effects included in complex models. A central point is that, although the models constructed through this procedure are `simple', in the sense that they have far fewer parameters than the models they are derived from and are more amenable to analysis, they will typically have features that would not have been guessed at if one started from simple models and then added further complexity. This is the strength of the approach: Model 1 contains within it a large amount of social science data and expertise, and a diluted form of this is retained in Model 3.

A direct translation of the methods used in physics would be to start with a minimal model, progressively introducing new structure and at every stage comparing the new model with data. We believe that the process we have described here is more directly suited to the social sciences, with its relative paucity of data. However the conventional physical sciences approach can still have a role after the various stages of models have been created. One could also attempt to go from Model $n$ to Model $(n-1)$, and in this way explore a wide range of possible models by going up and down the various stages. In this way one should be able to gain a fuller appreciation of the role that various extra structures have in giving a more complete description of the system.

To demonstrate the utility of creating a model that is amenable to mathematical analysis we have used Model 3 to investigate the origins of the bistability seen in Model 2, and the existence of high and low turnout regimes found in both Model 2 and Model 1. This investigation allowed us to understand the mechanisms required for bistability to exist, and provided an explanation for the observed dependence on the homogeneity of the population and the structure of the social network when one is used. This was a specific illustration of the use of this method in models of voter turnout, but we believe that the present approach of using a chain of increasingly simple models can be fruitful for the analysis of a wide variety of complex systems.

\medskip

This work was supported by the EPSRC under grant number EP/H02171X.

\medskip

\small{\textit{emails:}\ Luis F. Lafuerza \url{lflafuerza@gmail.com}; Louise Dyson \url{l.dyson@warwick.ac.uk}; Bruce Edmonds \url{bruce.edmonds@gmail.com}; Alan J. McKane \url{alan.mckane@manchester.ac.uk}}

\medskip

Author contribution statement: AJM and BE designed the research, LFL and LD performed the research, LFL, LD, AJM and BE wrote the paper.

\medskip

\appendix

\section{Description of Model 2}
\label{appendixmodel2}
The description of Model 2 can be found in \cite{SCIDVoterUS}; we give it here for completeness.

Each agent (with index $i = 1, \ldots, N$) has the following list of characteristics, some of which can change over time:
\begin{itemize}
\item[] \textbf{binary variables:} civic duty ($cd(i)$), turnout (in last election, $v(i)$), voting habit ($h(i)$), post-18 education ($ed(i)$)
\item[] \textbf{integer variables:} (political) interest level ($l(i)$), minimum interest level ($m(i)$), age (in years, $a(i)$), number of (political) conversations remembered ($c(i)$)
\end{itemize}

The main parameters of the model are:\newline
Influence rate $K$, which scales the number of (political) conversations per year.\newline
Probabilities of initiating a conversation $\mathbf{p}_c(l,v)$.\newline
Probabilities of gaining and losing civic duty $p_{acd}(e,v)$ and $p_{lcd}(e,v,a)$, respectively.\newline
Thresholds on the number of conversations needed to increase the interest level $T_{\alpha}$.\newline
Probability of forgetting a conversation $p_f(l)$.\newline
Death probability $p_d(a)$.\newline
Emigration probability $p_e$.\newline
Probability of not voting due to confounding factors $p_c(a)$.

\subsection{Initialisation procedures}\label{Initialisation}
Agents are initialised using data derived from the British Household Panel Study (BPS) \cite{BHPS}. The same procedure initialises immigrants into the model, using the subset of the BHPS corresponding to survey responses from immigrants. This procedure sets the civic duty, turnout, voting habit, post-18 education, interest level and minimum interest level, with some of these characteristics being inferred using proxies for the required information. Agents initially do not remember any conversations, and have an age drawn from a uniform distribution between 18 and 70 (to initialise the model) and between 18 and 48 (for later immigrants into the model). Agents born into the simulation have age 18 (they are only taken into account in the model when they are adults), and education with probability 0.3. Their interest level and minimum interest level is equal to their education, and they are assumed have no civic duty, voting habit or conversations remembered and not to have voted in the last election.

\subsection{Main loop}
The following processes happen in a loop until the required time-point is reached. All rates are given in Table~\ref{parametervalues} below.

\textbf{Each year:} 
	\begin{itemize}
		\item[] \textbf{Each month:}
		\begin{itemize}
			\item[] \textbf{Carrying out conversations:} For each agent, this section is run $\lfloor $K$ / 12 \rfloor$ times plus one time extra with probability $ $K$ / 12 - \lfloor $K$ / 12 \rfloor$. 
			\begin{itemize}
				\item[] The agent has the chance to initiate three conversations, with probabilities $\mathbf{p}_c(l(i),v(i))$  each with a random other agent.
				\item[] Agents (with $l(i)>0$) receiving a conversation (from an agent with civic duty), acquire civic duty with probability $p_{acd}(ed(i),v(i))$.
			\end{itemize} 
			\item[] \textbf{Updating interest levels:} 
			\begin{itemize}
				\item[] If $l(i)=0$ and $c(i)>T_0$ then set $l(i) = 1$ and $m(i) = 1$.
				\item[] Else, if $c(i)>T_h$ then set $l(i) = m(i) + 2$.
				\item[] Else, if $c(i)>T_l$ then set $l(i) = m(i) + 1$.
			\end{itemize}
			\item[] \textbf{Updating civic duty:} Agents lose civic duty with probability, $p_{lcd}(ed(i),v(i))$, dependent on their age and education.
		\end{itemize}
		\item[] \textbf{Forgetting conversations:} Agents forget conversations that happened more than one year ago, with probability, $p_f(l(i))$, per conversation, dependent on the agent's interest level.
		\item[] \textbf{Birth/death:} Each agent dies with a probability, $p_d(a(i))$, dependent on their age, and is replaced by a new agent by the `birth' process (described in Section \ref{Initialisation}).
		\item[] \textbf{Immigration/emigration:} Each agent emigrates with a probability $p_e = 0.015$ and is replaced by a new agent by the `immigration' process (described in Section \ref{Initialisation}).
		\item[] \textbf{Ageing:} Agents age by one year.
	\end{itemize}
\textbf{Every 5 years there is an election:}
	\begin{itemize}
		\item[] Agents with civic duty or voting habit vote unless `confounded' (due to illness or other factors) with probability $p_c(a(i))$, dependent on their age.
		\item[] Agents gain voting habit if they vote in 3 consecutive elections.
		\item[] Agents lose voting habit if they do not vote in 2 consecutive elections.
	\end{itemize}
Here $\lfloor x\rfloor$ denotes the integer part of $x$, that is, the largest integer no greater than $x$.

\begin{table*}[h!!]
\begin{tabular}{|l|l|l|}
\hline
\textbf{Parameter}         & \textbf{Value}                            & \textbf{Meaning}                                                                                                \\
\textbf{name} &&\\ \hline
N                               &  480                                      & population size                                                                                        \\ \hline
$p_d(a)$                        & a function of age derived from mortality tables	& death rate                                                                                             \\ \hline
$p_e$                           & 0.015                                     & emigration rate                                                                                        \\ \hline
$K$                             & $K \in [2,12]$                            & influence rate                                                                                         \\ \hline
\multirow{6}{*}{$\mathbf{p}_c(l,v)$}     & $\mathbf{p}_c(2,0) = [0.0100,0.0500,0.1500]$  & \multirow{6}{*}{probability of initiating a conversation ($\mathbf{p}_c(l,v) = [0,0,0]$ if $l\leq 1$)} \\
                                & $\mathbf{p}_c(2,1) = [0.0600,0.1000,0.1800]$  &                                                                                                        \\
                                & $\mathbf{p}_c(3,0) = [0.0600,0.1925,0.3795]$  &                                                                                                        \\
                                & $\mathbf{p}_c(3,1) = [0.1540,0.2800,0.3900]$  &                                                                                                        \\
                                & $\mathbf{p}_c(4,0) = [0.2000,0.4750,0.5134]$  &                                                                                                        \\
                                & $\mathbf{p}_c(4,1) = [0.3232,0.5680,0.5370]$  &                                                                                                        \\ \hline
$p_{acd}(e,v)$                    & 1-(1-0.25(1+e)(1+v))(1-0.010(1+e)(1+v)) & probability of acquiring civic duty (per conversation)                                                                   \\ \hline
$T_0$                          & 5                                         & threshold for increasing interest level to 1                                                           \\ \hline
$T_l$                          & 2                                         & lower threshold for increasing interest                                                                \\ \hline
$T_h$                          & 5                                         & higher threshold for increasing interest                                                               \\ \hline
\multirow{2}{*}{$p_{lcd}(a,e)$} & $0.01/(12(1+e))$ if $a\geq 25$           & \multirow{2}{*}{probability (per month) of losing civic duty}                                                      \\ \cline{2-2}
                                & $0$ if $a < 25$                           &                                                                                                        \\ \hline
\multirow{2}{*}{$p_c(a)$}       &  $0.077$ if $a\leq 75$                    & \multirow{2}{*}{probability of not voting due to being confounded}                        \\ \cline{2-2}
                                & $0.077+(1-0.077)0.9^{(a-75)(a-74)/2}$  else   &                                                                                                        \\ \hline
$p_f$       & 0.2                             & probability (per year) of forgetting a conversation                                                                                                                                                    \\ \hline
$p_{fb}$       & 0.5                             & probability (per year) of forgetting a background conversation                                                                                                                                                    \\ \hline
\end{tabular}
\caption{Parameter values for Model 2}
\label{parametervalues}
\end{table*}

\section{Parameters for Model 3}\label{appendixparam}
In the main text, Model 3 has been defined in rather general terms. In order to correspond to Model 2 as described in Appendix~\ref{appendixmodel2}, it is necessary to make the following identifications:

The social interaction function $f$ has to take the following piece-wise constant form: 
 \begin{equation}
 f(s,m)=\begin{cases}
    0, & \text{if $s<2c$}.\\
    0.322\equiv f_1, & \text{if $2c\leq s<3c$}.\\
    0.794\equiv f_2, & \text{if $3c\leq s<4c$}.\\
    0.794\equiv f_2, & \text{if $s\geq 4c$ and $m<2c$}.\\
    1.397\equiv f_3, & \text{if $s\geq 4c$ and $m\geq 2c$}.\label{fModel2}
\end{cases}
\end{equation}
These numbers are taken directly from Model 2, except for the following approximation. In Model 2 the $f(s,m)$ is affected by whether or not the agent voted in the previous election. This link is broken in Model 3. In order to have comparable values for $f$, we choose the ones corresponding to Model 2, assuming that the agent voted with a probability of 0.86. This is approximately equal to the proportion of agents voting in the high-communication regime (we expect it to work as well on the low-communication regime because the few agents that initiate conversations in that regime are also likely to have voted).

The probability of acquiring and losing civic duty is given by:
 \begin{equation}
 a_d=\begin{cases}
    0.258, & \text{if $v(i)=ed(i)=0$}.\\
    0.51, & \text{if $v(i)=1, ed(i)=0$ or $v(i)=0, ed(i)=1$}.\\
    1, & \text{if $ed(i)=v(i)=1$}.
\end{cases}
\end{equation}
 \begin{equation}
 l_d=\begin{cases}
    0.000417, & \text{if $ed(i)=0$}.\\
    0.00083, & \text{else}.\\
\end{cases}
\end{equation}
Other parameter values are $\gamma=0.0167$ month$^{-1}$, $T_d=c$, $p_c=0.139$, $\tau_e=5$ years. The total number of individuals is $N=500$. Also $c=3$ unless otherwise stated. \newline
Individuals born in the simulation are initiated with 18 years of age (we do not explicitly include children), with: $ed(i)=B(0.34), m(i)=c\cdot ed(i), v(i)=h(i)=d(i)=0$.\newline
The characteristics of immigrants are set based on statistics derived from the British Household Panel Study \cite{BHPS}, as follows:
$m(i)=3c, 2c, c, 0$, with probabilities $0.02, 0.06, 0.21, 0.71$, respectively.\newline
If $m(i)=3c$, then $s(i)=4c$, $ed(i)=d(i)=1$, $v(i)=B(0.9), h(i)=B(0.29)$.\newline
If $m(i)=2c$, then $s(i)=3c+B(0.32), ed(i)=B(0.68), d(i)=B(0.43), v(i)=B(0.70), h(i)=B(0.13)$.\newline
If $m(i)=c$, then $s(i)=c, ed(i)=1, d(i)=B(0.21), v(i)=0.72, h(i)=B(0.13)$.\newline
If $m(i)=0$, then $s(i)=c, ed(i)=0, d(i)=B(0.21), v(i)=0.72, h(i)=B(0.13)$.\newline
In addition, if $h(i)=v(i)=1$, with probability 0.9 it is assumed that the agent voted in the election previous to the latest one (this is relevant for the dynamics of voting habit). Here $B(x)$ denotes a Bernoulli random variable with mean $x$, that is $B(x)=1$ with probability $x$, $B(x)=0$ else. These parameter values are taken directly from Model 2, with no data fitting involved.

With the form of $f(s,m)$ given in (\ref{fModel2}), equation (\ref{self-consnice}) becomes:
\begin{eqnarray}
&&\frac{\gamma}{K}\lambda=P(0)f_2+P(1)f_2+P(2)f_3+P(3)f_3\nonumber\\
&&-\frac{\Gamma(c,\lambda)}{\Gamma(c)}\left[P(1)f_1+P(2)(f_2-f_1)+P(3)(f_3-f_2)\right]\nonumber\\
&&-\frac{\Gamma(2c,\lambda)}{\Gamma(2c)}[P(0)f_1+P(1)(f_2-f_3)+P(2)(f_3-f_2)]\nonumber\\
&&-\frac{\Gamma(3c,\lambda)}{\Gamma(3c)}P(0)(f_2-f_1),\label{selfconssicd}
\end{eqnarray}
where $\Gamma(a)=\Gamma(a,0)$ is the Gamma function, and $\Gamma(a,x)\equiv\int_x^{\infty}t^{a-1}e^{-t}dt$ is the incomplete Gamma function. Since $\Gamma(a,x)/\Gamma(a)$ decreases monotonically with $x$, with $\Gamma(a,0)/\Gamma(a)=1, \Gamma(a,\infty)/\Gamma(a)=0, \Gamma(a,a)/\Gamma(a)\approx1/2$ (last approximate equality being valid for large $a$), we see that the right-hand side of (\ref{selfconssicd}) changes from $P(2)f_1+P(3)f_2$ for $\lambda\lesssim1$ (when only agents with intrinsic state equal to two or three initiate conversations) to $P(0)f_2+P(1)f_2+P(2)f_3+P(3)f_3$ for $\lambda\gtrsim3c$ (when all agents initiate conversations at the maximum possible rate given their intrinsic state). The right hand side of (\ref{selfconssicd}) typically displays an S-shape which can lead to several solutions for $\lambda$ in a range of parameter values, as illustrated in Fig.~\ref{self-consfig} in the main text. 

For $\lambda\ll c$, Eq. (\ref{selfconssicd}) simplifies to:
\begin{equation}
 \frac{\gamma}{K}\lambda\simeq P(2)f_1+P(3)f_2\Rightarrow \gamma\lambda\simeq K[P(2)f_1+P(3)f_2],
\end{equation}
while for $\lambda\ll c$, Eq. (\ref{selfconssicd}) leads to:
\begin{eqnarray}
 &&\frac{\gamma}{K}\lambda\simeq [P(0)+P(1)]f_2+[P(2)+P(3)]f_3\\
&&\Rightarrow \gamma\lambda\simeq K\{[P(0)+P(1)]f_2+[P(2)+P(3)]f_3\}.\nonumber
\end{eqnarray}
We see that both the solution with small $\lambda$ and the one with large $\lambda$, increase linearly with $K$. This simple approximation breaks down for intermediate values of $\lambda$ (of the order of $c$), but it can be rather accurate, as evidenced by the approximately straight character of the theoretical lines in figure (\ref{convsfig}) of the main text.


\begin{thebibliography}{24}
\expandafter\ifx\csname natexlab\endcsname\relax\def\natexlab#1{#1}\fi
\expandafter\ifx\csname bibnamefont\endcsname\relax
  \def\bibnamefont#1{#1}\fi
\expandafter\ifx\csname bibfnamefont\endcsname\relax
  \def\bibfnamefont#1{#1}\fi
\expandafter\ifx\csname citenamefont\endcsname\relax
  \def\citenamefont#1{#1}\fi
\expandafter\ifx\csname url\endcsname\relax
  \def\url#1{\texttt{#1}}\fi
\expandafter\ifx\csname urlprefix\endcsname\relax\def\urlprefix{URL }\fi
\providecommand{\bibinfo}[2]{#2}
\providecommand{\eprint}[2][]{\url{#2}}

\bibitem[{\citenamefont{Zuckerman}(2005)}]{Zuckerman05}
\bibinfo{editor}{\bibfnamefont{A.~S.} \bibnamefont{Zuckerman}}, ed.,
  \emph{\bibinfo{title}{The Social Logic of Politics}}
  (\bibinfo{publisher}{Temple University Press},
  \bibinfo{address}{Philadelphia}, \bibinfo{year}{2005}).

\bibitem[{\citenamefont{Rolfe}(2012)}]{Rolfe12}
\bibinfo{author}{\bibfnamefont{M.}~\bibnamefont{Rolfe}},
  \emph{\bibinfo{title}{Voter Turnout: A Social Theory of Political
  Participation}} (\bibinfo{publisher}{Cambridge University Press},
  \bibinfo{address}{New York}, \bibinfo{year}{2012}).

\bibitem[{\citenamefont{Sinclair}(2012)}]{Sinclair12}
\bibinfo{author}{\bibfnamefont{B.}~\bibnamefont{Sinclair}},
  \emph{\bibinfo{title}{The Social Citizen: Peers Networks and Political
  Behavior}} (\bibinfo{publisher}{University of Chicago Press},
  \bibinfo{address}{Chicago}, \bibinfo{year}{2012}).

\bibitem[{\citenamefont{Nickerson}(2008)}]{Nickerson08}
\bibinfo{author}{\bibfnamefont{D.~W.} \bibnamefont{Nickerson}},
  \bibinfo{journal}{American Political Science Review}
  \textbf{\bibinfo{volume}{102}}, \bibinfo{pages}{49} (\bibinfo{year}{2008}).

\bibitem[{\citenamefont{Bond et~al.}(2012)\citenamefont{Bond, Fariss, Jones,
  Kramer, Marlow, Settle, and Fowler}}]{FBvoting12}
\bibinfo{author}{\bibfnamefont{R.~M.} \bibnamefont{Bond}},
  \bibinfo{author}{\bibfnamefont{C.~J.} \bibnamefont{Fariss}},
  \bibinfo{author}{\bibfnamefont{J.~J.} \bibnamefont{Jones}},
  \bibinfo{author}{\bibfnamefont{A.~D.~I.} \bibnamefont{Kramer}},
  \bibinfo{author}{\bibfnamefont{C.}~\bibnamefont{Marlow}},
  \bibinfo{author}{\bibfnamefont{J.~E.} \bibnamefont{Settle}},
  \bibnamefont{and} \bibinfo{author}{\bibfnamefont{J.~H.}
  \bibnamefont{Fowler}}, \bibinfo{journal}{Nature}
  \textbf{\bibinfo{volume}{489}}, \bibinfo{pages}{295} (\bibinfo{year}{2012}).

\bibitem[{\citenamefont{Goldenfeld and Kandanoff}(1999)}]{complexity}
\bibinfo{author}{\bibfnamefont{N.}~\bibnamefont{Goldenfeld}} \bibnamefont{and}
  \bibinfo{author}{\bibfnamefont{L.~P.} \bibnamefont{Kandanoff}},
  \bibinfo{journal}{Science} \textbf{\bibinfo{volume}{284}},
  \bibinfo{pages}{87} (\bibinfo{year}{1999}).

\bibitem[{\citenamefont{Bernardes et~al.}(2002)\citenamefont{Bernardes,
  Stauffer, and Kert{\'e}sz}}]{Sznajd1}
\bibinfo{author}{\bibfnamefont{A.~T.} \bibnamefont{Bernardes}},
  \bibinfo{author}{\bibfnamefont{D.}~\bibnamefont{Stauffer}}, \bibnamefont{and}
  \bibinfo{author}{\bibfnamefont{J.}~\bibnamefont{Kert{\'e}sz}},
  \bibinfo{journal}{Eur. Phys. J. B} \textbf{\bibinfo{volume}{25}},
  \bibinfo{pages}{123} (\bibinfo{year}{2002}).

\bibitem[{\citenamefont{Gonzalez et~al.}(2004)\citenamefont{Gonzalez, Sousa,
  and Herrmann}}]{Sznajd2}
\bibinfo{author}{\bibfnamefont{M.}~\bibnamefont{Gonzalez}},
  \bibinfo{author}{\bibfnamefont{A.}~\bibnamefont{Sousa}}, \bibnamefont{and}
  \bibinfo{author}{\bibfnamefont{H.}~\bibnamefont{Herrmann}},
  \bibinfo{journal}{Int. J. Mod. Phys. C} \textbf{\bibinfo{volume}{15}},
  \bibinfo{pages}{45} (\bibinfo{year}{2004}).

\bibitem[{\citenamefont{Borghesi and Bouchaud}(2010)}]{Borghesi1}
\bibinfo{author}{\bibfnamefont{C.}~\bibnamefont{Borghesi}} \bibnamefont{and}
  \bibinfo{author}{\bibfnamefont{J.-P.} \bibnamefont{Bouchaud}},
  \bibinfo{journal}{Eur. Phys. J. B} \textbf{\bibinfo{volume}{75}},
  \bibinfo{pages}{395} (\bibinfo{year}{2010}).

\bibitem[{\citenamefont{Borghesi et~al.}(2012)\citenamefont{Borghesi, Raynal,
  and Bouchaud}}]{Borghesi2}
\bibinfo{author}{\bibfnamefont{C.}~\bibnamefont{Borghesi}},
  \bibinfo{author}{\bibfnamefont{J.-C.} \bibnamefont{Raynal}},
  \bibnamefont{and} \bibinfo{author}{\bibfnamefont{J.-P.}
  \bibnamefont{Bouchaud}}, \bibinfo{journal}{PloS one}
  \textbf{\bibinfo{volume}{7}}, \bibinfo{pages}{e36289} (\bibinfo{year}{2012}).

\bibitem[{\citenamefont{Galam and Moscovici}(1991)}]{Galam91}
\bibinfo{author}{\bibfnamefont{S.}~\bibnamefont{Galam}} \bibnamefont{and}
  \bibinfo{author}{\bibfnamefont{S.}~\bibnamefont{Moscovici}},
  \bibinfo{journal}{Eur. J. Soc. Psychol.} \textbf{\bibinfo{volume}{21}},
  \bibinfo{pages}{49} (\bibinfo{year}{1991}).

\bibitem[{\citenamefont{Fowler}(2005)}]{Fowler}
\bibinfo{author}{\bibfnamefont{J.~H.} \bibnamefont{Fowler}}, in
  \emph{\bibinfo{booktitle}{The Social Logic of Politics}}, edited by
  \bibinfo{editor}{\bibfnamefont{A.~S.} \bibnamefont{Zuckerman}}
  (\bibinfo{publisher}{Temple University Press},
  \bibinfo{address}{Philadelphia}, \bibinfo{year}{2005}), pp.
  \bibinfo{pages}{296--287}.

\bibitem[{\citenamefont{Fosco et~al.}(2011)\citenamefont{Fosco, Laruelle, and
  S\'anchez}}]{Anxovoting11}
\bibinfo{author}{\bibfnamefont{C.}~\bibnamefont{Fosco}},
  \bibinfo{author}{\bibfnamefont{A.}~\bibnamefont{Laruelle}}, \bibnamefont{and}
  \bibinfo{author}{\bibfnamefont{A.}~\bibnamefont{S\'anchez}},
  \bibinfo{journal}{Advances in Complex Systems} \textbf{\bibinfo{volume}{14}},
  \bibinfo{pages}{31} (\bibinfo{year}{2011}).

\bibitem[{\citenamefont{Fern\'andez-Gracia
  et~al.}(2014)\citenamefont{Fern\'andez-Gracia, Suchecki, Ramasco, San~Miguel,
  and Egu\'{i}luz}}]{Voterdata}
\bibinfo{author}{\bibfnamefont{J.}~\bibnamefont{Fern\'andez-Gracia}},
  \bibinfo{author}{\bibfnamefont{K.}~\bibnamefont{Suchecki}},
  \bibinfo{author}{\bibfnamefont{J.~J.} \bibnamefont{Ramasco}},
  \bibinfo{author}{\bibfnamefont{M.}~\bibnamefont{San~Miguel}},
  \bibnamefont{and} \bibinfo{author}{\bibfnamefont{V.~M.}
  \bibnamefont{Egu\'{i}luz}}, \bibinfo{journal}{Phys. Rev. Lett.}
  \textbf{\bibinfo{volume}{112}}, \bibinfo{pages}{158701}
  (\bibinfo{year}{2014}).

\bibitem[{\citenamefont{Edmonds et~al.}(2014)\citenamefont{Edmonds,
  Lessard-Phillips, and Fieldhouse}}]{modelref}
\bibinfo{author}{\bibfnamefont{B.}~\bibnamefont{Edmonds}},
  \bibinfo{author}{\bibfnamefont{L.}~\bibnamefont{Lessard-Phillips}},
  \bibnamefont{and}
  \bibinfo{author}{\bibfnamefont{E.}~\bibnamefont{Fieldhouse}},
  \emph{\bibinfo{title}{A complex model of voter turnout (version 1)}},
  \bibinfo{howpublished}{CoMSES Computational Model Library.
  \url{https://www.openabm.org/model/4368/version/1}} (\bibinfo{year}{2014}).

\bibitem[{\citenamefont{Loughran et~al.}(2015)\citenamefont{Loughran,
  Lessard-Phillips, Fieldhouse, and Edmonds}}]{modelref2}
\bibinfo{author}{\bibfnamefont{T.}~\bibnamefont{Loughran}},
  \bibinfo{author}{\bibfnamefont{L.}~\bibnamefont{Lessard-Phillips}},
  \bibinfo{author}{\bibfnamefont{E.}~\bibnamefont{Fieldhouse}},
  \bibnamefont{and} \bibinfo{author}{\bibfnamefont{B.}~\bibnamefont{Edmonds}},
  \emph{\bibinfo{title}{The voter model - a long description}},
  \bibinfo{howpublished}{SCID project document, available at
  \url{https://scidproject.files.wordpress.com/2015/05/the-voter-model-description-final-draft.docx}}
  (\bibinfo{year}{2015}).

\bibitem[{\citenamefont{Lafuerza et~al.}()\citenamefont{Lafuerza, Dyson,
  Edmonds, and McKane}}]{SCIDVoterUS}
\bibinfo{author}{\bibfnamefont{L.~F.} \bibnamefont{Lafuerza}},
  \bibinfo{author}{\bibfnamefont{L.}~\bibnamefont{Dyson}},
  \bibinfo{author}{\bibfnamefont{B.}~\bibnamefont{Edmonds}}, \bibnamefont{and}
  \bibinfo{author}{\bibfnamefont{A.~J.} \bibnamefont{McKane}},
  \bibinfo{journal}{PLoS ONE} \textbf{\bibinfo{volume}{11}},
  \bibinfo{pages}{e0157261} (\bibinfo{year}{2016}).

\bibitem[{\citenamefont{Van~Kampen}(2007)}]{VanKampenBook}
\bibinfo{author}{\bibfnamefont{N.~G.} \bibnamefont{Van~Kampen}},
  \emph{\bibinfo{title}{Stochastic Processes in Physics and Chemistry}}
  (\bibinfo{publisher}{Elsevier}, \bibinfo{address}{Amsterdam},
  \bibinfo{year}{2007}), \bibinfo{note}{p. 145}.

\bibitem[{\citenamefont{Granovetter}(1978)}]{Granovetter78}
\bibinfo{author}{\bibfnamefont{M.}~\bibnamefont{Granovetter}},
  \bibinfo{journal}{Am. J. Sociol.} \textbf{\bibinfo{volume}{83}},
  \bibinfo{pages}{1420} (\bibinfo{year}{1978}).

\bibitem[{\citenamefont{Siegel}(2009)}]{Siegel09}
\bibinfo{author}{\bibfnamefont{D.~A.} \bibnamefont{Siegel}},
  \bibinfo{journal}{Am. J. Polit. Sci.} \textbf{\bibinfo{volume}{53}},
  \bibinfo{pages}{122} (\bibinfo{year}{2009}).

\bibitem[{\citenamefont{Sethna et~al.}(1993)\citenamefont{Sethna, Dahmen,
  Kartha, Krumhansl, Roberts, and Shore}}]{randomIsing}
\bibinfo{author}{\bibfnamefont{J.~P.} \bibnamefont{Sethna}},
  \bibinfo{author}{\bibfnamefont{K.}~\bibnamefont{Dahmen}},
  \bibinfo{author}{\bibfnamefont{S.}~\bibnamefont{Kartha}},
  \bibinfo{author}{\bibfnamefont{J.~A.} \bibnamefont{Krumhansl}},
  \bibinfo{author}{\bibfnamefont{B.~W.} \bibnamefont{Roberts}},
  \bibnamefont{and} \bibinfo{author}{\bibfnamefont{J.~D.} \bibnamefont{Shore}},
  \bibinfo{journal}{Phys. Rev. Lett.} \textbf{\bibinfo{volume}{70}},
  \bibinfo{pages}{3347} (\bibinfo{year}{1993}).

\bibitem[{\citenamefont{Michard and Bouchaud}(2005)}]{Bouchaud05}
\bibinfo{author}{\bibfnamefont{Q.}~\bibnamefont{Michard}} \bibnamefont{and}
  \bibinfo{author}{\bibfnamefont{J.-P.} \bibnamefont{Bouchaud}},
  \bibinfo{journal}{Eur. J. Soc. Psychol.} \textbf{\bibinfo{volume}{47}},
  \bibinfo{pages}{151} (\bibinfo{year}{2005}).

\bibitem[{\citenamefont{Bouchaud}(2013)}]{BouchaudReview}
\bibinfo{author}{\bibfnamefont{J.-P.} \bibnamefont{Bouchaud}},
  \bibinfo{journal}{J. Stat. Phys.} \textbf{\bibinfo{volume}{151}},
  \bibinfo{pages}{567} (\bibinfo{year}{2013}).

\bibitem[{BHP()}]{BHPS}
\emph{\bibinfo{title}{British household panel study ({BHPS})}},
  \bibinfo{howpublished}{\url{https://www.iser.essex.ac.uk/bhps}}.

\end{thebibliography}

\end{document}